\documentclass[runningheads]{llncs}
\usepackage[hyphens]{url}
\usepackage{graphicx}
\usepackage{acronym}
\usepackage{subfig}

\PassOptionsToPackage{utf8}{inputenc}
  \usepackage{inputenc}

\PassOptionsToPackage{T1}{fontenc} 
  \usepackage{fontenc}

\newcommand{\la}{LA\ }

\newcommand{\lsvp}{LSVP\ }

\AtBeginDocument{%
  \def\doi#1{\url{https://doi.org/#1}}}
\makeatother

\begin{document}
\title{A Survey of Left Atrial Appendage Segmentation and Analysis in 3D and 4D Medical Images
\thanks{Supported in part by the Croatian Science Foundation under project UIP-2017-05-4968.}
}

%
%
\author{
  Hrvoje Leventić\inst{1}\orcidID{0000-0001-9568-1408} \and
  Marin Benčević\inst{1}\orcidID{0000-0003-4294-0781} \and
  Danilo Babin\inst{2}\orcidID{0000-0002-2881-6760} \and
  Marija Habijan\inst{1}\orcidID{0000-0002-3754-498X} \and
  Irena Galić\inst{1}\orcidID{0000-0002-0211-4568}
}
\authorrunning{H. Leventić et al.}
%
\institute{
  J. J. Strossmayer University of Osijek, Faculty of Electrical
  Engineering, Computer Science and Information Technologies, Osijek, Croatia\\
  \email{\{hrvoje.leventic,marin.bencevic,marija.habijan,irena.galic\}@ferit.hr}\\
\and
TELIN-IPI, Faculty of Engineering and Architecture,\\
Ghent University – imec, Belgium\\
\email{danilo.babin@ugent.be}}
\maketitle              
\begin{abstract}
Atrial
fibrillation (AF) is a cardiovascular disease identified as one of the main risk
factors for stroke. The majority of strokes due to AF are
caused by clots originating in the left atrial appendage (LAA). LAA
occlusion is an effective procedure for reducing stroke risk. 
Planning the procedure using pre-procedural imaging and analysis has shown
benefits. The analysis is
commonly done by manually segmenting the appendage on 2D slices. Automatic LAA
segmentation methods could save an expert's time and provide insightful 3D
visualizations and accurate automatic measurements to aid in medical
procedures. Several semi- and fully-automatic methods for segmenting the appendage have
been proposed. This paper provides a review of automatic LAA segmentation
methods on 3D and 4D medical images, including CT, MRI and echocardiogram
images. We classify methods into heuristic and model-based methods, as well as
into semi- and fully-automatic methods. We summarize and compare the proposed
methods, evaluate their effectiveness, present current challenges in the field
and approaches to overcome them.

\end{abstract}

\section{Introduction}

Cardiovascular diseases (CVDs) have been identified as the leading cause of
death in the developed world, with stroke accounting for about a third of
all CVD deaths \cite{worldhealthorganization2017_top10causes}.  Atrial
fibrillation (AF) is a CVD identified as one of the main risk
factors for stroke
\cite{goldman1999_Pathophysiologiccorrelatesthromboembolism}.  Atrial
fibrillation presents as asynchronous chaotic contractions of the atria.
The atrium fibrillates,
i.e. contract outside the standard regular sinus rhythm of the heart. The blood
pools in the atria, enabling the formation of a blood clot, which can cause a
stroke if it becomes dislodged and enters the bloodstream. The majority of 
strokes due to AF are caused by clots
originating in the left atrial appendage (LAA)
\cite{cabrera2014_Leftatrialappendage}.
While the prevention of stroke due to AF is commonly done with 
anticoagulation therapy (i.e. warfarin), patients often have
contraindications to this type of therapy. 
Left atrial appendage occlusion (LAAO) is an alternative stroke prevention
procedure which avoids most of the drawbacks of anticoagulation therapy. 
During the procedure, a device is percutaneously deployed into the
neck of the LAA which prevents the blood flow to the appendage, and stops the
blood clot from exiting the appendage and entering the circulatory system.

Multiple occluder devices are available on the market. Device choice is
dependent on the patient's LAA anatomy, as it can vary significantly between
patients. Wang et al. \cite{wang2010_LeftAtrialAppendage} classify LAA
morphology into four types: 
chicken-wing (48\% of patients), cactus (30\%), windsock (19\%) and cauliflower (3\%
-- most often associated with embolic events).
Choosing the correct
device for the patient's anatomy requires
accurate measurements of the heart and the appendage. The measurements can be
obtained using medical imaging such as transesophageal echocardiography (TEE),
standard fluoroscopy and computed tomography (CT).

While the procedure can be performed using standard fluoroscopy and TEE,
 using a pre-procedural CT imaging to plan the procedure
and to guide the sizing of the device has shown benefits, including better
prediction of the appropriate device size
\cite{saw2016_ComparingMeasurementsCT,wang2016_Application3DimensionalComputed},
and better determination of patient's suitability for the procedure
\cite{wang2016_Application3DimensionalComputed}. Often,
physicians perform the measurements directly in 2D slices of different
multi-planar reconstruction (MPR) views. Accurate image processing methods for
segmentation and analysis of the LAA can aid physicians in reducing the time to plan the
occlusion procedure, by calculating the required measurements and visualizing
the appendage using an accurate 3D model.
Accurate LAA segmentation methods can also be used in a workflow for diagnosing
atrial fibrillation \cite{jin2018_DetectionSubstancesLeft}, or predicting
if the patient is at an increased risk of thrombus formation
\cite{bosi2018_ComputationalFluidDynamic}. It can also provide more accurate
measurements of other clinically important parameters like LAA emptying
velocity \cite{bosi2018_ComputationalFluidDynamic}.

The goal of this paper is to provide a review of the current state-of-the-art
methods in LAA segmentation and analysis in 3D and 4D medical images. The paper
is organized as follows.
Section \ref{sec:segmentation} provides an overview of methods which focus on
LAA segmentation from 3D and 4D medical images. Section \ref{sec:cfd} provides
an overview of methods which focus on predicting the risk of atrial
fibrillation development or thrombus formation. A discussion about the current
state of the research is given in Section \ref{sec:discussion}. Finally,
Section \ref{sec:conclusion} gives a conclusion.

\section{LAA segmentation methods}
\label{sec:segmentation}

This section covers the state-of-the-art methods for the segmentation of the
left atrial appendage. Segmentation approaches presented in this section can be divided into two segments: heuristics-based
methods (section \ref{sec:segmentation_heu}) and model-based methods
(section \ref{sec:segmentation_ml}). Presented approaches can also
be divided according to the degree of user interaction required: into fully
automatic methods and semi-automatic methods. 
LAA segmentation is still often performed manually as well,
using guided region-growing-based segmentation methods 
\cite{garcia-isla2018_Sensitivityanalysisgeometrical} and software such as 3D Slicer
\cite{hecko2020_Data3Dleft}.
However, this paper is focused on approaches designed specifically for LAA segmentation.

\subsection{Heuristics-based segmentation methods}
\label{sec:segmentation_heu}

This section presents heuristics-based LAA segmentation methods. 
Most of the methods described in this section are based on a modification of
region growing. Likewise, most of the methods are semi-automatic and require
user interaction. Some methods require one or more user-selected
seed points \cite{vdovjak2019_AdaptiveThresholdingSingle,leventic2019_Leftatrialappendage}, 
thresholds
\cite{leventic2019_Leftatrialappendage}, 
or a user provided centerline through the LAA
\cite{morais2018_FastSegmentationLeft}.

Morais et al. \cite{morais2018_FastSegmentationLeft} proposed a centerline
based LAA segmentation approach which works in 3D TEE images. The approach initializes
a model from a manually created centerline, grows the model using fast contour growing and 
determines  the segmentation from the refined model. The method is evaluated on
20 TEE datasets manually segmented by two observers. 
Metrics used to evaluate the method were:
point-to-surface error (P2S), Dice similarity and 95th percentile Hausdorff
distance. Obtained results were close to the inter- and intra-observer
variability. Dice similarity score was around 82.5\%. The method runtime is
under 20 seconds with an Intel i7 CPU at 2.8 GHz.

\begin{figure}[t]
  \centering
  \includegraphics[width=\textwidth]{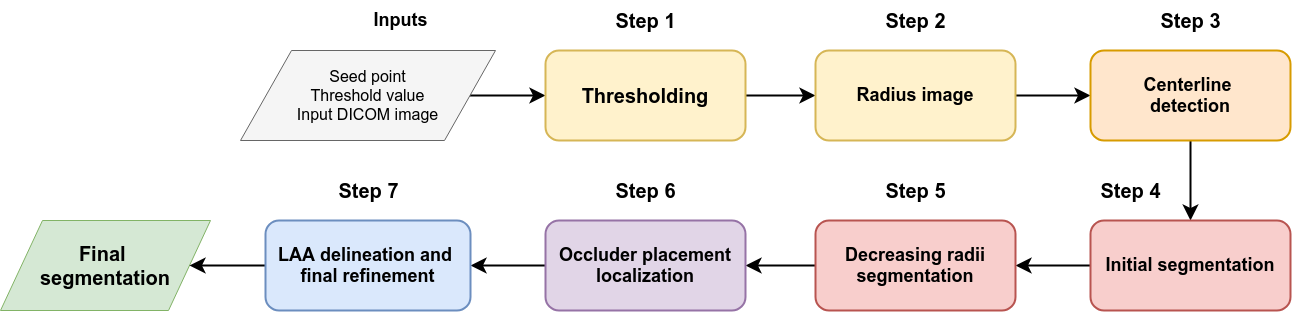}
  \caption[]{
    The LAA segmentation framework flow diagram by Leventic et al. 
    \cite{leventic2019_Leftatrialappendage}.  
  }
  \label{fig:laaseg_proposed_seg_flow}
\end{figure}

Our own proposed semi-automatic LAA segmentation framework is presented in  
\cite{leventic2019_Leftatrialappendage,vdovjak2019_AdaptiveThresholdingSingle}
and illustrated in Figure \ref{fig:laaseg_proposed_seg_flow}.
The framework requires a single seed point (single click) inside the LAA. An input CT image is first
thresholded using an adaptive thresholding method to produce a binary mask
image. Radius image is calculated from the mask image using Euclidean distance
transform (EDT) and used to extract a centerline connecting the
seed point to the center of the left atrium. The method
reconstructs an approximate LAA volume from the centerline by combining the largest maximum
inscribed spheres of every point in the centerline. Then, a novel
decreasing radii segmentation algorithm is used to refine the segmentation. Finally, 
the localization of the LAA orifice is performed by finding a plane delineating the LAA from the left
atrium, thus proposing a location for the occluder placement. The method is
invariant to the type and dimensions of the input image used for segmentation,
as it can work with binary images created from any type of 3D image (CT,
MRI, etc.), so long as it contains the LAA and
at least a part of the left atrium. 
The method is validated against 17 CT datasets manually segmented by two
medical experts. Dice similarity score is 91.54\% and 87.93\% (against
the two expert segmentations) when using automatic threshold detection, or
92.52\% and 91.53\% when selecting the threshold value manually.

Moghadam A. et al. \cite{A2020} propose a semi-automatic method to segment the
LAA device landing zone from echocardiogram images for the purpose of LAA
occlusion procedures. Optimal 2D axial slices are selected by an expert for
each echo dataset. LAA appears as an ellipsoidal dark region,
which their algorithm aims to detect. Each slice is first shifted to the same
region of interest. The slice is then smoothed using a Gaussian filter, gamma corrected and
sharpened. The image is thresholded and binarized, producing an image with
several black holes. 8-adjacent black pixels (pixels for which
each neighboring pixel has the same value) represent the segmentation. 
The area, perimeter, dimensions and position of the segmented
regions is used to refine the segmentation of the LAA landing
zone. The authors report correct segmentation for 18 out of 22 (81.8\%) echo datasets,
the algorithm failed to obtain a segmentation for the remaining cases. The authors
do not perform validation of the accuracy of the successful segmentations.

Jia et al. \cite{jia2019_ImageBasedFlowSimulations} proposed an automatic LAA
segmentation method for the purpose of computational fluid dynamics (CFD) simulation. 
The method automatically
detects a seed point in the LA by detecting the ascending and descending aorta
in an axial view and utilizing the fact that the LA is located between them.
The segmentation is performed using a Bayesian inference region-growing method
and the detected seed point. Several parameters in the method
are determined either empirically, or from the knowledge of the heart anatomy. 
The runtime of the method is between 0.5-1 min and it achieves a Dice
similarity score of 86.3\%. However, the method is only validated on 5 ground
truth images.

\subsection{Model-based segmentation methods}
\label{sec:segmentation_ml}

This section presents model-based LAA segmentation methods. These include
machine and deep learning based approaches, as well as other methods that
include learned models like atlas-based segmentation. Until recently, only a few fully
automatic segmentation methods were published in the literature. However, the
advancement deep neural networks for segmentation resulted in a several new
fully automatic segmentation approaches being proposed. 

Atlas-based segmentation utilizes atlases, images that are labeled maps of
structures that need to be segmented \cite{Rohlfing2005}. Atlases are usually obtained via manual
segmentation by an expert. Those atlases can then be used to segment new unseen
images through a process called registration. Registration transforms and
deforms an input image such that the structures on the image map onto the same
structures of a target image. Mathematical scoring functions are used
to determine how well one image maps onto the other. In atlas-based
segmentation, input images are registered to one or more atlas images. Once
registered, the segmentation is performed by finding the corresponding atlas
label on the registered image for each pixel of the input image. Atlas-based
segmentation requires fewer annotated images than machine-learning-based
methods \cite{Qiao2019,zheng2008_FourChamberHeartModelinga}.

Usually atlas-based segmentation is done by using multiple atlas images. The
input image is registered onto each atlas separately, and the segmentation
results are then fused using different mathematical criteria. This approach
often suffers from a "diminishing distal part" problem: Segmentation is
worse at the edges of an object than at the center.

Qiao et al. \cite{Qiao2019} present a fully automatic method to segment the \la
chamber, pulmonary veins and LAA from MRA (magnetic resonance angiography)
images from the MICCAI'13 LASC challenge
\cite{tobon-gomez2015_BenchmarkAlgorithmsSegmenting}. They use atlas-based
segmentation which aims to reduce the diminishing distal part problem. They
register all atlases to the image in a single step by
formulating the registration as a group objective function optimization
process. Their algorithm optimizes the variance between registrations of each
atlas, thus jointly registering the input image to all atlases without losing
accuracy at the edges of the segmented regions. The method runtime is 
around 9 minutes per scan. They use 10 MRA scans for training, and another
20 for testing. The reported Dice similarity score for LAA segmentation is 91\%.

Another common model-based approach for image segmentation is machine learning.
Machine learning for image segmentation is usually done by extracting a
selection of features from images. These features often include pixel gray
levels, pixel locations, image moments, information about a pixel's
neighborhood, etc.  A vector of image features is then fed into a learned
classifier which classifies each pixel of the image into a class.  The
parameters of the classifier are learned automatically by giving the classifier
input images for which the ground truth classification results is known. The
output of the model can then be compared to the ground truth, and the
parameters of the model are adjusted so that the model's output better matches
the ground truth value.  This procedure is repeated for a large amount of input
images, so that the learned parameters generalize to new, unseen
examples. The process of adjusting the model's parameters is called training.

Deep learning, a subset of machine learning, eliminates the need for
pre-programmed feature extraction. Instead, the feature extraction itself is
learned by the model during training.  Deep-learning based image segmentation
is commonly done using convolutional neural networks (CNNs).  Convolutional
neural networks have a layered structure where series of convolutions are
performed on an input image. Kernels of the convolutions are learned during
training. The convolution results are then combined using a learned statistical
model that outputs a segmented image. 

The first two fully automatic methods for LAA segmentation were proposed by Zheng et al.
\cite{zheng2008_FourChamberHeartModelinga,zheng2014_MultiPartModelingSegmentation}
and evaluated in the LASC challenge
\cite{tobon-gomez2015_BenchmarkAlgorithmsSegmenting}. The methods are referred
to as SIE-PMB and SIE-MRG respectivelly in the challenge.  Zheng et al.
(SIE-PMB) \cite{zheng2008_FourChamberHeartModelinga} uses a multi-part based
model fitting approach to automatically segment the left atrium, including the
LAA and the pulmonary veins. Each of the six parts of the model is
fitted individually using marginal space learning (MSL) and later merged into a consolidated
mesh.
This method was trained on an in-house CT dataset of 457 cardiac CT scans, one of the largest training datasets
reported in the literature.  Main limitation of the method is the size of the
dataset required for training, while the main strength is the computational
efficiency: the complete \la segmentation (including the LAA) in 3 seconds on a
multi-core CPU. 
The second approach (SIE-MRG) \cite{zheng2014_MultiPartModelingSegmentation} uses a
similar multi-part based approach.
The difference is that segmentation refinement using region-growing based with
adaptive thresholds is performed after model fitting, 
followed by graph-cuts-based 
removal of the leakage. The region-growing-based refinement allows
better capturing of the varying morphology of the LAA, as well as the
trabeculations inside the LAA. 

\begin{figure}[t]
  
  \includegraphics[width=\linewidth]{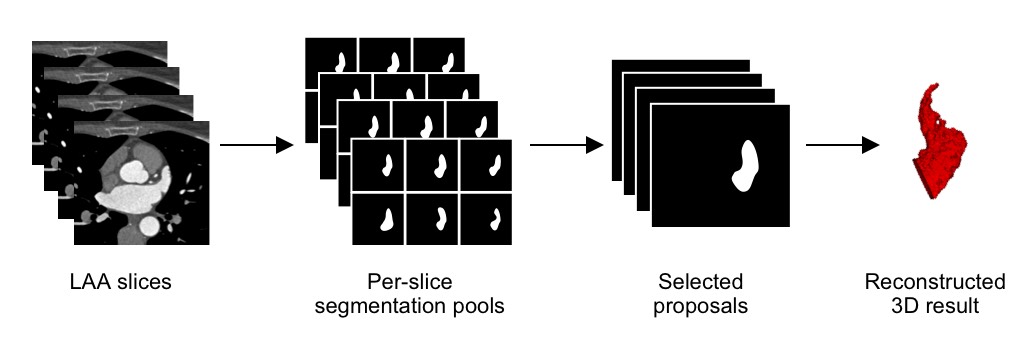}
  \caption{
    Flowchart of the LAA segmentation method proposed by Wang et al. \cite{wang2016_LeftAtrialAppendage}.
  }
  \label{fig:bg_sota_laa_wang_flow}  
\end{figure}

\begin{figure}[t]
  
  \includegraphics[width=\linewidth]{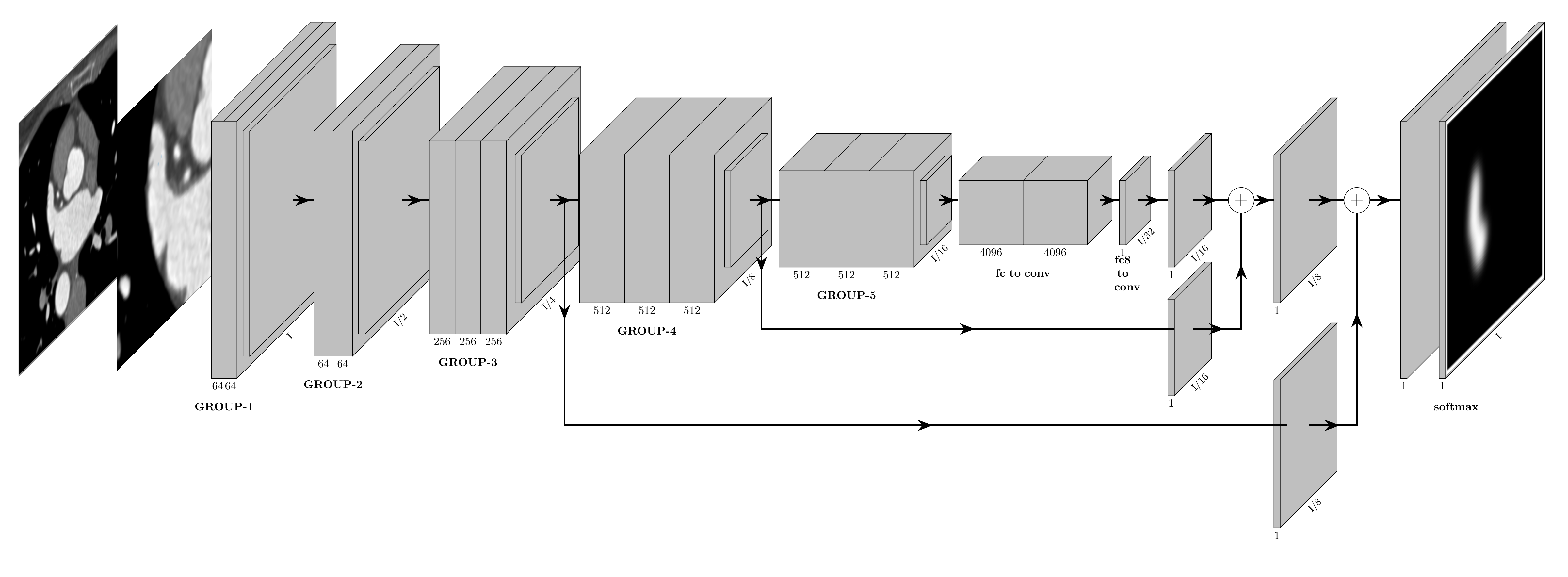}
  \caption{
    Details of the network architecture proposed by Jin et al. \cite{jin2018_LeftAtrialAppendage}.
    The network combines abstract, high-level
    semantic features with low-level spatial information.
  }
  \label{fig:bg_sota_laa_jin_cnn}  
\end{figure}

A semi-automatic LAA segmentation method was proposed by Wang et al.
\cite{wang2016_LeftAtrialAppendage}. The method segments the LAA from 3D
cardiac CTA images using an approach based on ranking 2D segmentation
proposals.  The flowchart of the proposed method is shown in Figure
\ref{fig:bg_sota_laa_wang_flow}.  Initally, the method requires manual
determination of the LAA bounding box.  Next, the  method processes all slices
of the determined LAA volume and for each slice creates a pool of segmentation
proposals.  A trained random forest regressor picks the best segmentation
proposal for each slice. Finally, best proposals are merged into a 3D volume
using spatial continuity to correct possible segmentation errors.
The method was evaluated on 60 CTA datasets and achieved a Dice score of
95.12\%. The method takes about 3.5 minutes on a 4 CPU system (4
Intel Core i7 CPUs at 4.0GHz) to perform the segmentation.

Jin et al. \cite{jin2017_LeftAtrialAppendage} proposed a method for LAA neck
modelling to aid in occlusion procedure planning.  The method segments the LAA
using \cite{wang2016_LeftAtrialAppendage}, automatically detects the LAA
ostium, and calculates the neck dimensions.  Ostium is detected as a smooth
closed boundary of the highest surface curvature located in the transitional
region between the \la and the appendage.  Finally, the method  models the
tension of the LAA surface after the closure procedure.  The method is
evaluated on 100 CT datasets and 3 pig hearts, with the post-procedural
follow-up of 67 patients indicating the 97.01\% success rate of occluder device
implantation (only two failed implantation cases). 

Jin et al. also proposed an LAA segmentation method \cite{jin2018_LeftAtrialAppendage}
based on fully
convolutional neural networks and conditional random fields. The method is an
improvement of \cite{wang2016_LeftAtrialAppendage} and uses a similar approach:
LAA is segmented in each 2D slice of the manually provided
bounding box.
Slices are pre-processed and converted to
3-channel RGB pseudo color images to enhance the resolution of local features.
Pseudo color images are input to the CNN
(architecture in Figure \ref{fig:bg_sota_laa_jin_cnn}) which outputs 2D
probability maps of the regions containing the appendage.  The final
segmentation step merges the 2D probability maps using 3D conditional
random fields into a final 3D volume. 
Training and evaluation is performed on 150 CCTA datasets using five-fold cross-validation. 
Achieved Dice overlap with the ground truth is 94.76\%. The runtime on
a single LAA volume is around 35 seconds on a Tesla K80 GPU.

Grasland-Mongrain et al.
\cite{grasland-mongrain2010_Combinationshapeconstrainedinflation}
propose an adaptation of active shape models (ASM)
\cite{ecabert2008_Automaticmodelbasedsegmentation} for LAA segmentation.
The heart is localized and segmented with shape-constrained deformable
models (ASM-based approach from
\cite{ecabert2008_Automaticmodelbasedsegmentation}).  The model-based approach
segments the heart chambers and determines the position of the appendage.
The localized appendage (the part of the mesh denoting LA-LAA interface)
is inflated in order to obtain the segmentation, using the minimization of the
internal and external energy.
The external energy pushes the mesh towards the appendage edges,
while the internal energy preserves a regular triangle distribution during the
inflation.
The method has
been evaluated on images from 17 patients.  The inflation of the mesh has
problems reaching the tip of the appendage, as well as undersegmenting the
appendage. The method is also sensitive to the selection of parameters for the
segmentation.

Al et al. \cite{al2018_ActorCriticReinforcementLearning}  proposed an automatic
LAA segmentation method using actor-critic reinforcement learning. First, 
actor-critic agents localize LAA and \lsvp seed points. Seed expansion
following surface trend using EDT provides a starting
segmentation and creates a volume-of-interest (VOI). 
Finally, the method classifies each voxel in VOI into LAA class and \lsvp
class, depending on which seed is closer. The method is evaluated on 28
annotated volumes and achieves 93.64\% Dice similarity score, with the runtime
of around 8 seconds per dataset.

\section{4D LAA analysis methods and CFD}
\label{sec:cfd}

\subsubsection{4D LAA analysis methods}

Jin et al. proposed two 4D-based LAA analysis methods
\cite{jin2018_Leftatrialappendagea,jin2018_DetectionSubstancesLeft}.  The 
method in \cite{jin2018_Leftatrialappendagea} performs the segmentation of 4D CT
LAA images to assist AF diagnosis. A 3D model of each time instance of
the sequence is built using their graph-cuts based segmentation approach
\cite{wang2016_LeftAtrialAppendage}.  
The method assists in AF diagnosis by:
(1) calculating the volume of 3D models in different phases of the cardiac cycle;
(2) generating the "volume-phase" curve (showing the change of LAA volume throughout the 
cycle); and (3) obtaining important dynamic LAA metrics. Finally, multivariate logistic regression 
analysis of the obtained metrics calculates the risk of thrombus formation, while 
the SVM-based model predicts the AF diagnosis.

The method in \cite{jin2018_DetectionSubstancesLeft} detects the substances
inside the LAA from 4D CT images using spatio-temporal motion analysis.
The method extracts the optical flow field for all adjacent phases in a cardiac cycle.
The cardiac cycle of 20 phases results in 19 optical flow fields. 
The method generates the motion trajectory of the key voxels
throughout the cycle using nearest neighbour interpolation.
Hierarchical clustering tree finds the corresponding classification for every trajectory 
track. Changes in classifications between the tracks correspond to the division of 
substances in the appendage. Finally, time-frequency analysis of the trajectories
enables the detection of different substances inside the appendage, including the 
thrombi in different states of formation. 

\subsubsection{LAA analysis using CFD}
Computational fluid dynamics (CFD) can be used for LAA analysis by calculating
important hemodynamic parameters such as LAA emptying velocity and visualizing
the flow of blood through the LAA. Bosi et al.
\cite{bosi2018_ComputationalFluidDynamic} estimated blood residence time in the
LAA in four typical LAA morphologies using virtual contrast agent washing
out, while imposing both healthy and AF conditions. While TEE can be used to
measure blood flow velocity in the LAA, it can only provide average information
on the velocity values. CFD approaches can visualize the regions where blood
pools in the LAA for longer than a single cardiac cycle. They had to manually
segment the LA and the LAA to perform the CFD simulations. 

Other works
\cite{otani2016_ComputationalFrameworkPersonalized,markl2016_LeftAtrialLeft} 
also proposed a
computational approach for personalized blood flow analysis from patient's CT
images. However, they also perform manual segmentation of the LAA and heart
chambers. 
Similar work was proposed by 
Grigoriadis et al. \cite{grigoriadis2020_ComputationalFluidDynamics}. They
used CFD to simulate LAA blood flow for 3 patients, using semi-automatically
segmented CT images. Their simulations showed
that blood velocity and wall shear stress of the LAA are decreased along the
LAA, thus making the tip of the LAA more prone to fluid stagnation.

Jia et al. \cite{jia2019_ImageBasedFlowSimulations} proposed a simulation
framework which, using patient specific CT images, evaluates the clinical
impact of the LAA, as well as the efficacy of the LAA closure. The framework
implements an automatic LAA segmentation method, described in
\ref{sec:segmentation_heu}. They perform the simulations in both pre- and
post-closure images and the framework successfuly predicts patient-specific
outcome of LAA closure. However, the CFD analysis is only performed for one
patient.

\section{Discussion }
\label{sec:discussion}
We have divided the presented methods into two
general categories: heuristics-based methods and model-based methods.
Presented methods could also be categorized by the amount of user interaction
required for successful segmentation. Until recently there were only a few
fully automatic methods in the literature. Lately, a number of fully automatic
methods has been published. Two main advantages of automatic methods are:
reducing the subjectivity of the segmentation and saving the expert's time.
However, the advantage of semi-automatic methods is in providing the expert the
control over the segmentation process and the ability to correct the
segmentation if errors occur, while still reducing the time required to perform
the segmentation. 

With the advancements in the computing power, blood flow simulations using
computational fluid dynamics show great potential for personalized medicine.
Some of the potential uses of CDF simulations are non-invasive prediction of
the risk of thrombus formation, AF diagnosis, prediction of the success of the
occlusion procedure, etc. However, the availability of fast and accurate
segmentation algorithms could greatly advance the proliferation of the CDF
simulation use in everyday clinical practice. 

Nonetheless, comparing different segmentation approaches is quite challenging.  
To the best of our knowledge, there are no public LAA segmentation datasets
available.  
Almost all presented approaches use in-house datasets. Datasets often contain
low number of images and ground truth segmentations are created by in-house
medical experts. However, several papers
\cite{leventic2019_Leftatrialappendage,morais2018_FastSegmentationLeft} have
shown non-negligible subjectivity between the ground truth segmentations from
different medical experts. The approaches that quantify and highlight this
subjectivity are mostly heuristics-based, since these approaches do not require
definite ground truth segmentations for training.  Approaches based on machine-
and deep-learning require a single \emph{correctly} labeled ground truth for
every image in the dataset. Thus, such approaches use a consensus of
multiple medical experts in creating the ground truth labels
\cite{jin2018_LeftAtrialAppendage,wang2016_LeftAtrialAppendage}.

\section{Conclusion}
\label{sec:conclusion}

This paper presents a review of the state-of-the-art methods for LAA
segmentation and analysis. We categorized the methods according to training
type and required user interaction, presented the methods and discussed
their strengths and limitations. Our analysis demonstrated that 
the last few years have shown great advancement in the area of LAA
segmentation and analysis. However, there is still room for improvement. Solving the
problem of accurate and fast LAA segmentation could increase the level of care
provided to AF and CVD patients, through advancing other uses such as CFD blood
flow simulations.

\bibliographystyle{splncs04}
\bibliography{zoterotest,marin}

\begin{thebibliography}{10}
\providecommand{\url}[1]{\texttt{#1}}
\providecommand{\urlprefix}{URL }
\providecommand{\doi}[1]{https://doi.org/#1}

\bibitem{A2020}
A, P.M.: A semi-automated algorithm for segmentation of the left atrial
  appendage landing zone: Application in left atrial appendage occlusion
  procedures. Journal of Biomedical Physics and Engineering  \textbf{10}(2)
  (Apr 2020). \doi{10.31661/jbpe.v0i0.1912-1019},
  \url{https://doi.org/10.31661/jbpe.v0i0.1912-1019}

\bibitem{al2018_ActorCriticReinforcementLearning}
Al, W.A., Yun, I.D.: Actor-{{Critic Reinforcement Learning}} for {{Automatic
  Left Atrial Appendage Segmentation}}. In: 2018 {{IEEE International
  Conference}} on {{Bioinformatics}} and {{Biomedicine}} ({{BIBM}}). pp.
  609--612 (Dec 2018). \doi{10.1109/BIBM.2018.8621575}

\bibitem{bosi2018_ComputationalFluidDynamic}
Bosi, G.M., Cook, A., Rai, R., Menezes, L.J., Schievano, S., Torii, R.,
  Burriesci, G.: Computational {{Fluid Dynamic Analysis}} of the {{Left Atrial
  Appendage}} to {{Predict Thrombosis Risk}}. Front Cardiovasc Med  \textbf{5}
  (Apr 2018). \doi{10.3389/fcvm.2018.00034}

\bibitem{cabrera2014_Leftatrialappendage}
Cabrera, J.A., Saremi, F., {S{\'a}nchez-Quintana}, D.: Left atrial appendage:
  Anatomy and imaging landmarks pertinent to percutaneous transcatheter
  occlusion. Heart  \textbf{100}(20),  1636--1650 (Oct 2014).
  \doi{10.1136/heartjnl-2013-304464}

\bibitem{ecabert2008_Automaticmodelbasedsegmentation}
Ecabert, O., Peters, J., Schramm, H., Lorenz, C., Von~Berg, J., Walker, M.J.,
  Vembar, M., Olszewski, M.E., Subramanyan, K., Lavi, G., et~al.: Automatic
  model-based segmentation of the heart in {{CT}} images. Med. Imaging IEEE
  Trans. On  \textbf{27}(9),  1189--1201 (2008),
  \url{http://ieeexplore.ieee.org/xpls/abs_all.jsp?arnumber=4505365}

\bibitem{garcia-isla2018_Sensitivityanalysisgeometrical}
Garc{\'i}a-Isla, G., Olivares, A.L., Silva, E., Nu{\~n}ez-Garcia, M., Butakoff,
  C., Sanchez-Quintana, D., Morales, H.G., Freixa, X., Noailly, J., Potter,
  T.D., Camara, O.: Sensitivity analysis of geometrical parameters to study
  haemodynamics and thrombus formation in the left atrial appendage. Int. J.
  Numer. Methods Biomed. Eng.  \textbf{34}(8),  e3100 (2018).
  \doi{10.1002/cnm.3100}

\bibitem{goldman1999_Pathophysiologiccorrelatesthromboembolism}
Goldman, M.E., Pearce, L.A., Hart, R.G., Zabalgoitia, M., Asinger, R.W.,
  Safford, R., Halperin, J.L., Investigators, S.P.i.A.F., et~al.:
  Pathophysiologic correlates of thromboembolism in nonvalvular atrial
  fibrillation: {{I}}. {{Reduced}} flow velocity in the left atrial appendage
  ({{The Stroke Prevention}} in {{Atrial Fibrillation}} [{{SPAF}}-{{III}}]
  study). J. Am. Soc. Echocardiogr.  \textbf{12}(12),  1080--1087 (1999),
  \url{http://www.sciencedirect.com/science/article/pii/S0894731799701057}

\bibitem{grasland-mongrain2010_Combinationshapeconstrainedinflation}
{Grasland-Mongrain}, P., Peters, J., Ecabert, O.: Combination of
  shape-constrained and inflation deformable models with application to the
  segmentation of the left atrial appendage. In: Biomedical {{Imaging}}: {{From
  Nano}} to {{Macro}}, 2010 {{IEEE International Symposium}} On. pp. 428--431.
  {IEEE} (2010), \url{http://ieeexplore.ieee.org/abstract/document/5490319/}

\bibitem{grigoriadis2020_ComputationalFluidDynamics}
Grigoriadis, G.I., Sakellarios, A.I., Naka, K., Kosmidou, I., Ellis, C.,
  Michalis, L.K., Fotiadis, D.I.: Computational {{Fluid Dynamics}} of {{Blood
  Flow}} at the {{Left Atrium}} and {{Left Atrium Appendage}}. In: Henriques,
  J., Neves, N., {de Carvalho}, P. (eds.) {{XV Mediterranean Conference}} on
  {{Medical}} and {{Biological Engineering}} and {{Computing}} \textendash{}
  {{MEDICON}} 2019. pp. 938--946. {{IFMBE Proceedings}}, {Springer
  International Publishing}, {Cham} (2020). \doi{10.1007/978-3-030-31635-8_114}

\bibitem{hecko2020_Data3Dleft}
Hecko, J., Jiravsky, O., Chovancik, J., Hudec, M., Sramko, M., Sknouril, L.:
  Data for {{3D}} left atrial printing acquired using open source and free
  software, with the aim to determine the proper size of left atrial appendage
  occlured. European Heart Journal  \textbf{41}(ehaa946.0665) (Nov 2020).
  \doi{10.1093/ehjci/ehaa946.0665}

\bibitem{jia2019_ImageBasedFlowSimulations}
Jia, D., Jeon, B., Park, H.B., Chang, H.J., Zhang, L.T.: Image-{{Based Flow
  Simulations}} of {{Pre}}- and {{Post}}-left {{Atrial Appendage Closure}} in
  the {{Left Atrium}}. Cardiovasc Eng Tech  \textbf{10}(2),  225--241 (Jun
  2019). \doi{10.1007/s13239-019-00412-7}

\bibitem{jin2018_LeftAtrialAppendage}
Jin, C., Feng, J., Wang, L., Liu, J., Yu, H., Lu, J., Zhou, J.: Left {{Atrial
  Appendage Segmentation Using Fully Convolutional Neural Networks}} and
  {{Modified Three}}-dimensional {{Conditional Random Fields}}. IEEE J. Biomed.
  Health Inform.  \textbf{PP}(99), ~1--1 (2018).
  \doi{10.1109/JBHI.2018.2794552}

\bibitem{jin2018_Leftatrialappendagea}
Jin, C., Feng, J., Wang, L., Yu, H., Liu, J., Lu, J., Zhou, J.: Left atrial
  appendage segmentation and quantitative assisted diagnosis of atrial
  fibrillation based on fusion of temporal-spatial information. Comput. Biol.
  Med.  \textbf{96},  52--68 (May 2018). \doi{10.1016/j.compbiomed.2018.03.002}

\bibitem{jin2017_LeftAtrialAppendage}
Jin, C., Yu, H., Feng, J., Wang, L., Lu, J., Zhou, J.: Left {{Atrial Appendage
  Neck Modeling}} for {{Closure Surgery}}. In: Statistical {{Atlases}} and
  {{Computational Models}} of the {{Heart}}. {{ACDC}} and {{MMWHS Challenges}}.
  pp. 32--41. Lecture {{Notes}} in {{Computer Science}}, {Springer, Cham} (Sep
  2017). \doi{10.1007/978-3-319-75541-0_4}

\bibitem{jin2018_DetectionSubstancesLeft}
Jin, C., Yu, H., Feng, J., Wang, L., Lu, J., Zhou, J.: Detection of
  {{Substances}} in the {{Left Atrial Appendage}} by {{Spatiotemporal Motion
  Analysis Based}} on {{4D}}-{{CT}}. In: Pop, M., Sermesant, M., Jodoin, P.M.,
  Lalande, A., Zhuang, X., Yang, G., Young, A., Bernard, O. (eds.) Statistical
  {{Atlases}} and {{Computational Models}} of the {{Heart}}. {{ACDC}} and
  {{MMWHS Challenges}}. pp. 42--50. Lecture {{Notes}} in {{Computer Science}},
  {Springer International Publishing} (2018)

\bibitem{leventic2019_Leftatrialappendage}
Leventi{\'c}, H., Babin, D., Velicki, L., Devos, D., Gali{\'c}, I., Zlokolica,
  V., Romi{\'c}, K., Pi{\v z}urica, A.: Left atrial appendage segmentation from
  {{3D CCTA}} images for occluder placement procedure. Computers in Biology and
  Medicine  \textbf{104},  163--174 (Jan 2019).
  \doi{10.1016/j.compbiomed.2018.11.006}

\bibitem{markl2016_LeftAtrialLeft}
Markl, M., Lee, D.C., Furiasse, N., Carr, M., Foucar, C., Ng, J., Carr, J.,
  Goldberger, J.J.: Left {{Atrial}} and {{Left Atrial Appendage 4D Blood Flow
  Dynamics}} in {{Atrial Fibrillation}}. Circ Cardiovasc Imaging
  \textbf{9}(9),  e004984 (Sep 2016). \doi{10.1161/CIRCIMAGING.116.004984}

\bibitem{morais2018_FastSegmentationLeft}
Morais, P., Queir{\'o}s, S., Meester, P.D., Budts, W., Vila{\c c}a, J.L.,
  Tavares, J.M.R.S., D'hooge, J.: Fast {{Segmentation}} of the {{Left Atrial
  Appendage}} in {{3D Transesophageal Echocardiographic Images}}. IEEE Trans.
  Ultrason. Ferroelectr. Freq. Control pp.~1--1 (2018).
  \doi{10.1109/TUFFC.2018.2872816}

\bibitem{otani2016_ComputationalFrameworkPersonalized}
Otani, T., {Al-Issa}, A., Pourmorteza, A., McVeigh, E.R., Wada, S., Ashikaga,
  H.: A {{Computational Framework}} for {{Personalized Blood Flow Analysis}} in
  the {{Human Left Atrium}}. Ann Biomed Eng  \textbf{44}(11),  3284--3294 (Nov
  2016). \doi{10.1007/s10439-016-1590-x}

\bibitem{Qiao2019}
Qiao, M., Wang, Y., Berendsen, F.F., van~der Geest, R.J., Tao, Q.: Fully
  automated segmentation of the left atrium, pulmonary veins, and left atrial
  appendage from magnetic resonance angiography by joint-atlas-optimization.
  Medical Physics  \textbf{46}(5),  2074--2084 (Mar 2019).
  \doi{10.1002/mp.13475}, \url{https://doi.org/10.1002/mp.13475}

\bibitem{Rohlfing2005}
Rohlfing, T., Brandt, R., Menzel, R., Russakoff, D.B., Maurer, C.R.: Quo vadis,
  atlas-based segmentation? In: Handbook of Biomedical Image Analysis, pp.
  435--486. Springer {US} (2005). \doi{10.1007/0-306-48608-3_11},
  \url{https://doi.org/10.1007/0-306-48608-3_11}

\bibitem{saw2016_ComparingMeasurementsCT}
Saw, J., Fahmy, P., Spencer, R., Prakash, R., Mclaughlin, P., Nicolaou, S.,
  Tsang, M.: Comparing {{Measurements}} of {{CT Angiography}}, {{TEE}}, and
  {{Fluoroscopy}} of the {{Left Atrial Appendage}} for {{Percutaneous
  Closure}}. J Cardiovasc Electrophysiol  \textbf{27}(4),  414--422 (Apr 2016).
  \doi{10.1111/jce.12909}

\bibitem{tobon-gomez2015_BenchmarkAlgorithmsSegmenting}
{Tobon-Gomez}, C., Geers, A.J., Peters, J., Weese, J., Pinto, K., Karim, R.,
  Ammar, M., Daoudi, A., Margeta, J., Sandoval, Z., Stender, B., Zheng, Y.,
  Zuluaga, M.A., Betancur, J., Ayache, N., Chikh, M.A., Dillenseger, J., Kelm,
  B.M., Mahmoudi, S., Ourselin, S., Schlaefer, A., Schaeffter, T., Razavi, R.,
  Rhode, K.S.: Benchmark for {{Algorithms Segmenting}} the {{Left Atrium From
  3D CT}} and {{MRI Datasets}}. IEEE Trans. Med. Imaging  \textbf{34}(7),
  1460--1473 (Jul 2015). \doi{10.1109/TMI.2015.2398818}

\bibitem{vdovjak2019_AdaptiveThresholdingSingle}
Vdovjak, K., Leventi{\'c}, H., Habijan, M., Gali{\'c}, I.: Adaptive
  {{Thresholding}} for {{Single Click Left Atrial Appendage Segmentation}}. In:
  2019 {{International Symposium ELMAR}}. pp. 35--38 (Sep 2019).
  \doi{10.1109/ELMAR.2019.8918651}

\bibitem{wang2016_Application3DimensionalComputed}
Wang, D.D., Eng, M., Kupsky, D., Myers, E., Forbes, M., Rahman, M., Zaidan, M.,
  Parikh, S., Wyman, J., Pantelic, M., Song, T., Nadig, J., Karabon, P.,
  Greenbaum, A., O'Neill, W.: Application of 3-{{Dimensional Computed
  Tomographic Image Guidance}} to {{WATCHMAN Implantation}} and {{Impact}}
  on~{{Early Operator Learning Curve}}: {{Single}}-{{Center Experience}}. JACC:
  Cardiovascular Interventions  \textbf{9}(22),  2329--2340 (Nov 2016).
  \doi{10.1016/j.jcin.2016.07.038}

\bibitem{wang2016_LeftAtrialAppendage}
Wang, L., Feng, J., Jin, C., Lu, J., Zhou, J.: Left {{Atrial Appendage
  Segmentation Based}} on {{Ranking}} 2-{{D Segmentation Proposals}}. In:
  Statistical {{Atlases}} and {{Computational Models}} of the {{Heart}}.
  {{Imaging}} and {{Modelling Challenges}}. pp. 21--29. {Springer, Cham} (Oct
  2016). \doi{10.1007/978-3-319-52718-5_3}

\bibitem{wang2010_LeftAtrialAppendage}
Wang, Y., Di~Biase, L., Horton, R.P., Nguyen, T., Morhanty, P., Natale, A.:
  Left {{Atrial Appendage Studied}} by {{Computed Tomography}} to {{Help
  Planning}} for {{Appendage Closure Device Placement}}. J. Cardiovasc.
  Electrophysiol.  \textbf{21}(9),  973--982 (Sep 2010).
  \doi{10.1111/j.1540-8167.2010.01814.x}

\bibitem{worldhealthorganization2017_top10causes}
{World Health Organization}: The top 10 causes of death worldwide fact sheet.
  Tech. rep. (2017), \url{http://www.who.int/mediacentre/factsheets/fs310/en/}

\bibitem{zheng2008_FourChamberHeartModelinga}
Zheng, Y., Barbu, A., Georgescu, B., Scheuering, M., Comaniciu, D.:
  Four-{{Chamber Heart Modeling}} and {{Automatic Segmentation}} for 3-{{D
  Cardiac CT Volumes Using Marginal Space Learning}} and {{Steerable
  Features}}. IEEE Trans. Med. Imaging  \textbf{27}(11),  1668--1681 (Nov
  2008). \doi{10.1109/TMI.2008.2004421}

\bibitem{zheng2014_MultiPartModelingSegmentation}
Zheng, Y., Yang, D., John, M., Comaniciu, D.: Multi-{{Part Modeling}} and
  {{Segmentation}} of {{Left Atrium}} in {{C}}-{{Arm CT}} for
  {{Image}}-{{Guided Ablation}} of {{Atrial Fibrillation}}. IEEE Trans. Med.
  Imaging  \textbf{33}(2),  318--331 (2014). \doi{10.1109/TMI.2013.2284382}

\end{thebibliography}
\end{document}